\documentclass[nomathfonts]{aipproc}
\layoutstyle{8s}
\usepackage{amssymb,amsmath,amsthm,amscd,texdraw}
\graphicspath{{figures/}}
\makeindex

\title[Why MaxEnt?]
      {Why Maximum Entropy? A Non-axiomatic Approach\footnote{M. Grendar and M. Grendar, Jr.,
"Why Maximum Entropy? A Non-axiomatic Approach", in {\it Bayesian Inference and
Maximum Entropy Methods in Science and Engineering: 21-st International Workshop},
edited by R. L. Fry, pp. 375-379, American Institute of Physics, Melville,
NY, 2002, vol. CP617}}
\author{Marian Grendar}
{
 address = {marian.grendar@bb.telecom.sk}
,email = marian.grendar@bb.telecom.sk
}
\author{Marian Grendar, Jr.}
{
 address = {Institute of Measurement Science, Slovak Academy of Sciences\linebreak
           D\'ubravsk\'a cesta 9, Bratislava, 842 19, Slovakia. umergren@savba.sk}
,email = umergren@savba.sk
}

\makeatletter

\def\vc{\mathbf}

\newtheorem{theorem}{Theorem}

\newtheorem{defn}{Definition}
\newtheorem{note}{Note}
\newenvironment{pf}{{\sl Proof.\ }}{$\quad\Box$
\par\medskip\noindent\ignorespaces}
\makeatletter

\begin{document}
\begin{abstract}
Ill-posed inverse problems of the form $\vc y = \vc X\vc p$ where
$\vc y$ is $J$-dimensional vector of a data, $\vc p$ is
$m$-dimensional probability vector which can not be measured
directly and matrix $\vc X$ of observable variables is a known
$J\times m$ matrix, $J < m$, are frequently solved by Shannon's
entropy maximization (MaxEnt, ME). Several axiomatizations were
proposed (see for instance \cite{Shannon},  \cite{Khinchin},
\cite{Shore}, \cite{TTL}, \cite{Skilling}, \cite{Csiszar},  \cite{PV},
\cite{Garrett}, as well as \cite{Uffink} for a critique of some of
them) to justify the MaxEnt method (also) in this context. The main aim
of the presented work is two-fold: 1) to view the concept of
complementarity of MaxEnt and Maximum Likelihood (ML) tasks introduced
at \cite{GG2000} from a geometric perspective, and consequently
2) to provide  an intuitive and non-axiomatic answer to the 'Why
MaxEnt?' question.
\end{abstract}

\maketitle

\index{MaxEnt}
\index{Maximum Likelihood}
\index{complementarity}
\index{simplicity}

\section{Introduction}

The concept of complementarity of maximum entropy and
maximum likelihood tasks, proposed at \cite{GG2000}, is in
this vignette interpreted from a geometrical point of view of vectors.
Two key notions are introduced: collinearity
and coherence. In addition to shaping the complementarity into an
elegant form the collinearity/coherence concepts offer an
elementary answer to the persistent 'Why MaxEnt?' question.

\section{Collinearity and coherence: ML and MaxEnt}

\begin{defn}
System of events $\{A_1, A_2, \dots, A_m\}$ is equivalently
described by its distribution $\vc{p} = [p_1, p_2, \dots, p_m]$, or by its
potential $\vc u = [u_1, u_2, \dots, u_m]$. The relationship of
equivalence is
$$
\vc p = dist(\vc u) = \frac{1}{\sum_{i=1}^m e^{-u_i}} [e^{-u_1},
e^{-u_2}, \dots, e^{-u_m}]
$$
\end{defn}

\begin{theorem}
$dist(\vc u) = dist(\vc u + C)$, where C is any constant.
\end{theorem}

\begin{note}
The theorem implies that pmf is determined (induced) uniquely by potential,
and potential is determined by pmf up to an additive constant.
\end{note}

\begin{defn}
Mean value of potential $\vc u$ weighted by pmf $\vc p$ is the
scalar product of the vectors $\vc{up}$.
\end{defn}

\begin{defn}
Two {\it potential vectors\/} $\vc a$, $k\vc a$, where $k\in\mathbb{R}$, are
told to be collinear.
\end{defn}

\begin{defn}
Coherence of two pmfs $\vc p$ and $\vc q$ on potential $\vc u$
(or: with respect to potential $\vc u$, or: relative to potential $\vc u$)
is defined as
$$
coher_{\vc u} (\vc p, \vc q) = \vc{up} - \vc{uq}
$$

(Zero value means the maximal possible coherence, the greater the
(absolute) value, the smaller the coherence.)
\end{defn}

\begin{defn}
If coherence of two pmfs is zero, the pmfs are called coherent.
\end{defn}

\begin{theorem}
{\rm (ML task with simple exponential form)} Let $\vc X$ be a random sample
with vector of frequencies $\vc r$. Let $\vc{u}$ be a potential.
Then in class of distributions $dist(\lambda\vc{u})$ induced by collinear with
$\vc{u}$ potentials $\lambda\vc{u}$ the most likely one $dist(\lambda_0\vc{u})$ is
coherent with $\vc{r}$ relative to potential $\vc{u}$:
\begin{equation*}
coher_{\vc{u}}(\vc{r}, dist(\lambda_0\vc{u})) = 0
\end{equation*}
\end{theorem}
\begin{pf}
For a proof see \cite{GG2000}, proof of the Theorem 1.
\end{pf}

\begin{defn}
The mean value of potential $\vc u$ weighted by its own distribution is
called entropy of the potential,
$$
ent(\vc u) =  \vc u \, dist(\vc u)
$$
\end{defn}

\begin{theorem}
{\rm (MaxEnt task with simple potential)} Let $\vc X$ be a random sample
with vector of frequencies $\vc{r}$.
Let $\vc{u}$ be a potential. Then in class of
distributions $dist(\vc{v})$ which are coherent with $\vc{r}$
relative to potential $\vc{u}$
$$
coher_{\vc u}(\vc r, dist(\vc v)) = 0
$$
the most entropic one $dist(\vc{v}_0)$ is induced by potential $\vc{v}_0 =
\lambda_0\vc{u}$  collinear with $\vc{u}$.
\end{theorem}
\begin{pf}
For a proof see \cite{GG2000}, proof of the Theorem 1.
\end{pf}

\begin{figure}[!ht]
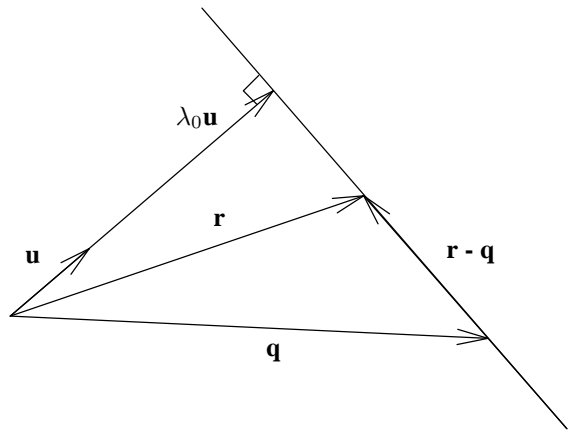

\begin{texdraw}
  \drawdim cm \linewd 0.02
  \arrowheadtype t:V
  \move(1 1.5)
  \ravec(6.35 -0.3)
  \move(0 0)
  \htext(4.4 0.9){\bf q}
  \move(1 1.5)
  \ravec(3.5 3)
  \move(0 0)
  \htext(3.2 4){$\lambda_0$\bf u}
  \move(1 1.5)
  \ravec(1.05 0.9)
  \move(0 0)
  \htext(1.2 2.2){\bf u}
  \move(8.40 0)
  \lvec(3.55 5.6)
  \move(1 1.5)
  \ravec(4.7 1.6)
  \move(0 0)
  \htext(3.7 2.7){\bf r}
  \move(8.4 0)
  \avec(5.7 3.1)
  \move(0 0)
  \htext(6.8 2.2){\bf{r} - \bf{q}}
  \move(4.3 4.7)
  \rlvec(-0.2 -0.2)
  \rlvec(0.18 -0.18)
\end{texdraw}
\caption{Geometric representation of Theorem 2, 3}
\end{figure}

From a geometrical standpoint, claims of the Theorem 1 and 2 lead to a search for
an intersection of $\vc u$ with a line defined by an orthogonal to $\vc u$ vector
$\vc{r} - \vc{q}$, as it is depicted on the Figure 1.

The circular relationship of the claims of Theorem 2 and Theorem
3, dubbed in \cite{GG2000} 'complementarity of Maximum Likelihood
and MaxEnt tasks', is visualized by the following diagram:

$$
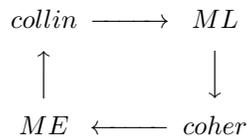
\begin{CD}
collin@>>> ML\\
@AAA@VVV \\
ME@<<<coher
\end{CD}
$$

In an intentionally loose manner the complementarity can be stated
as: 'in the class of collinear potentials the most probable (the most likely)
is the coherent one; in the class of coherent distributions the most
entropic is the collinear one'.

In light of the MaxProb rationale of MaxEnt (see \cite{GG}), the
above statement can achieve an even deeper symmetry, in the case
of sufficiently large sample. Then the words 'the most entropic' can
be replaced by 'the most probable', and the diagram can be
redrawn:

$$
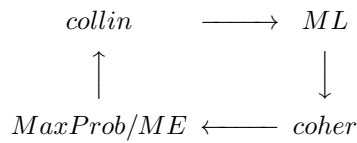
\begin{CD}
collin@>>> ML\\
@AAA@VVV \\
MaxProb/ME@<<<coher
\end{CD}
$$

\medskip\medskip

\section{Why MaxEnt? A simple answer}

Theorem 3 offers a simple argument in favor of
Shannon's entropy maximization (MaxEnt) method/criterion in the inverse-problem context,
which was mentioned at the abstract.

Consider the following 'game' which models the inverse problem:
an experiment reveals $m$ different
outcomes. The experiment was repeated sufficiently many times and vector of
frequencies of the outcomes in the obtained random sample $\vc r =
[r_1, r_2, \dots, r_m]$ is known to us. Also, we are told a
potential $\vc u$. Given this information $\{\vc r, \vc u\}$ we are asked
to pick up a potential $\vc v_0$ of a set of all potentials $\vc v$
whose distributions are coherent with $\vc r$ on potential $\vc u$
(in other words: the set consists of all such potentials
$\vc v$ that $\vc{ru} = dist(\vc v)\vc u$\,). Common sense dictates
to choose $\vc v_0 = \vc u$, but since it is unlikely that this
choice will satisfy the coherence condition, the second simplest
possible choice is a collinear with $\vc u$ potential $\vc v_0 = \lambda\vc
u$ where $\lambda$ should be chosen such that $dist(\lambda\vc u)$ will attain coherence
with $\vc r$ on the potential $\vc u$.

The fact that MaxEnt (and to the best of our knowledge no other method)
chooses in the above 'game' just the collinear
with $\vc u$ potential, hence {\it the simplest possible solution\/},
can be used as an answer to 'Why MaxEnt?' question.

In order to make relationship of the above mentioned 'game' to the ill-posed inverse
problem clear, note that the pair of information $\{\vc r, \vc u\}$ forms the value
of $y$, as $y = \vc r \vc u$, and the ill-posed problem becomes $y = \vc u \vc p$.
This way obviously extends also to more dimensional $\vc y$, hence to the case
where more potentials are given.

\section{Acknowledgments}
It is a pleasure to thank  Ale\v s Gottvald, George Judge, Gejza Wimmer and
Viktor Witkovsk\'y for valuable discussions and/or comments on earlier
version of this paper. The thanks extend also to participants of MaxEnt
workshop, especially to Robert Fry. The work was in part supported by the
grant VEGA 1/7295/20 from the Scientific Grant Agency of the Slovak
Republic. The second of the authors gratefully acknowledges financial
support of his participation at MaxEnt workshop by Jaynes' foundation.

\smallskip

Note. As compared to the paper which has appeared at the MaxEnt
proceedings, here Theorems 2 and 3 were re-stated in terms of random sample,
so that  the  argument made at the last Section should be easier to grasp.


\begin{thebibliography}{99}
\bibitem
{Shannon}
C. E. Shannon, ``A Mathematical Theory of Communication,'', Bell
System Technical Journal, {\bf 27}, pp. 379--423 and 623--656,
1948.
\bibitem
{Khinchin}
A. I. Khinchin, {\it Mathematical Foundations of Information Theory\/},
Dover, NY, 1957.
\bibitem
{Shore}
J. E. Shore and R. W. Johnson, ``Axiomatic Derivation of the
Principle of Maximum Entropy and the Principle of Minimum
Cross-Entropy,'' IEEE Transaction on Information Theory, {\bf IT-26},
pp. 26--37, 1980.
\bibitem{TTL}
Y. Tikoshinsky, N. Z. Tishby and R. D. Levine, ``Alternative
Approach to Maximum Entropy Inference,'' Physical Review A {\bf
30}, pp. 2638--2644, 1984
\bibitem
{Skilling}
J. Skilling, ``Classical Maximum Entropy,'' in {\it Maximum
Entropy and Bayesian Methods\/}, edited by J. Skilling, Dordrecht,
Kluwer, pp. 45--52, 1989.
\bibitem
{Csiszar}
I. Csisz\'ar, ``Why Least Squares and Maximum Entropy? An
Axiomatic Approach to Inference for Linear Inverse Problems,'',
Ann. Statist., {\bf 19}, pp. 2032-2056, 1991.
\bibitem
{PV}
J. Paris  and A. Vencovsk\'a, ``In Defense of Maximum Entropy
Inference Process,'' {\em Internat. J. Approx.
Reason.}, \textbf{17}, pp. 77-103, 1997.
\bibitem
{Garrett}
A. J. M. Garrett, ``Maximum Entropy from the Laws of Probability,'' in {\it
Bayesian Inference and Maximum Entropy Methods in Science and Engineering\/},
edited by A. Mohammad-Djafari, pp. 3--23, AIP Press, New York,  2001.
\bibitem{Uffink}
J. Uffink, ``Can the Maximum Entropy Principle Be Explained as a Consistency Requirement?,''
{\em Studies in History and Philosophy of modern Physics}, \textbf{26B}, pp. 223-261, 1995.
\bibitem
{GG2000}
M. Grend\'ar, Jr., and M. Grend\'ar, ``MiniMax Entropy and Maximum
Likelihood: Complementarity of Tasks, Identity of Solutions,'' in {\it
Bayesian Inference and Maximum Entropy Methods in Science and Engineering\/},
edited by A. Mohammad-Djafari, pp. 49--61, AIP Press, New York,  2001.
Available on-line at http://xxx.lanl.gov/abs/math.PR/0009129.
\bibitem
{GG}
M. Grend\'ar, Jr., and M. Grend\'ar, ``What Is the Question that MaxEnt
Answers? A Probabilistic Interpretation,'' in {\it Bayesian Inference and Maximum
Entropy Methods\/}, edited by A. Mohammad-Djafari, pp 83--94, AIP
Press, New York,  2001. Available on-line at
http://xxx.lanl.gov/abs/math-ph/009020.
\end{thebibliography}
\end{document}